\documentclass[a4paper,11pt]{article}
\usepackage{pos}

\usepackage{graphicx}
\usepackage{amsmath}
\usepackage{xspace}
\usepackage[numbers,sort&compress]{natbib}
\usepackage{subfig}
\usepackage[shortlabels]{enumitem}

\usepackage{pstricks}
\usepackage{graphicx}
\usepackage{xspace}
\usepackage{hyperref}
\usepackage{lipsum}
\usepackage{multirow}

\newcommand{\bea}{\begin{eqnarray}}
\newcommand{\eea}{\end{eqnarray}}

\newcommand{\lb}{\left(}
\newcommand{\rb}{\right)}
\newcommand{\eps}{\epsilon}

\newcommand{\ord}{\mathcal O}

\newcommand{\calI}{\mathcal{I}}

\newcommand{\gsim}{\;\rlap{\lower 3.5 pt \hbox{$\mathchar \sim$}} \raise 1pt
 \hbox {$>$}\;}
\newcommand{\lsim}{\;\rlap{\lower 3.5 pt \hbox{$\mathchar \sim$}} \raise 1pt
 \hbox {$<$}\;}

\title{
\vskip-3cm{\baselineskip14pt
    \begin{flushright}
     \normalsize \normalfont{CERN-TH-2025-195}
    \end{flushright}} \vskip2.5cm
Massive Feynman integrals at high energies: recent analytic results}

\author*[a]{Hantian Zhang}

\affiliation[a]{Theoretical Physics Department, CERN,\\
  1211 Geneva 23, Switzerland}

\emailAdd{hantian.zhang@cern.ch}

\abstract{
The high-energy behaviour of  scattering amplitudes involving massive particles has attracted interest in recent years.
In these proceedings, we report on the analytic tool \texttt{AsyInt}~\cite{Zhang:2024fcu} for solving massive multi-loop Feynman integrals in the high-energy limit,
which are fundamental building blocks for such amplitudes in the full Standard Model.
We present recent analytic results for two-loop four-point Feynman integrals with both internal and external masses in this limit,
featuring polylogarithmic and elliptic structures.
}

\FullConference{The European Physical Society Conference on High Energy Physics (EPS-HEP2025)\\
7-11 July 2025\\
Marseille, France\\}

\begin{document}
\maketitle

\allowdisplaybreaks

\begin{figure}[htb!]
  \centering
  \begin{tabular}{cc}
  \includegraphics[width=.25\textwidth]{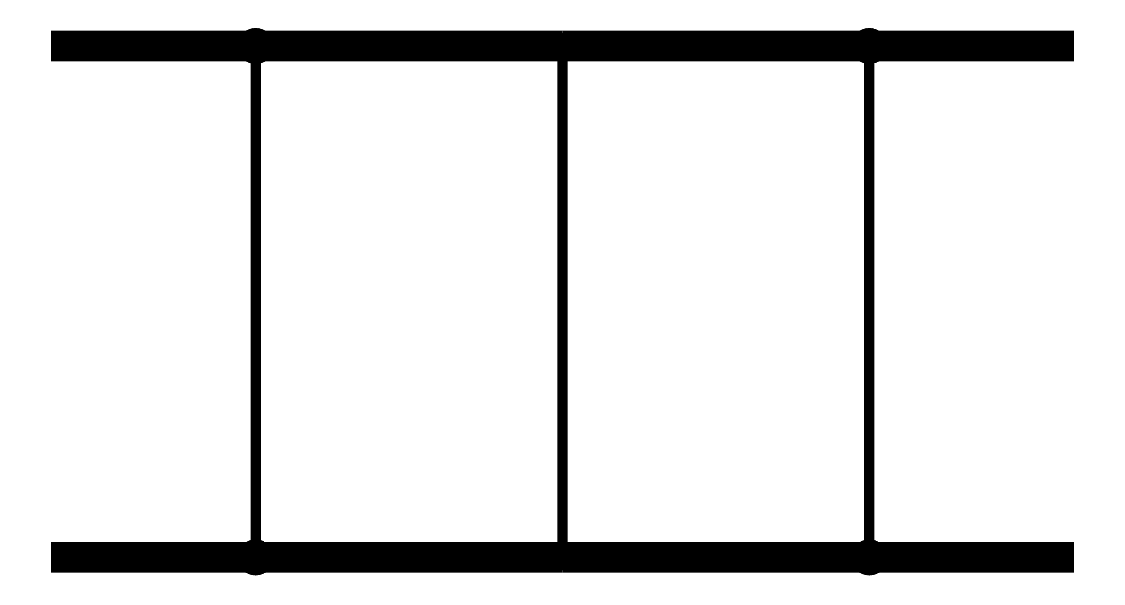}  \qquad & \qquad 
    \includegraphics[width=.25\textwidth]{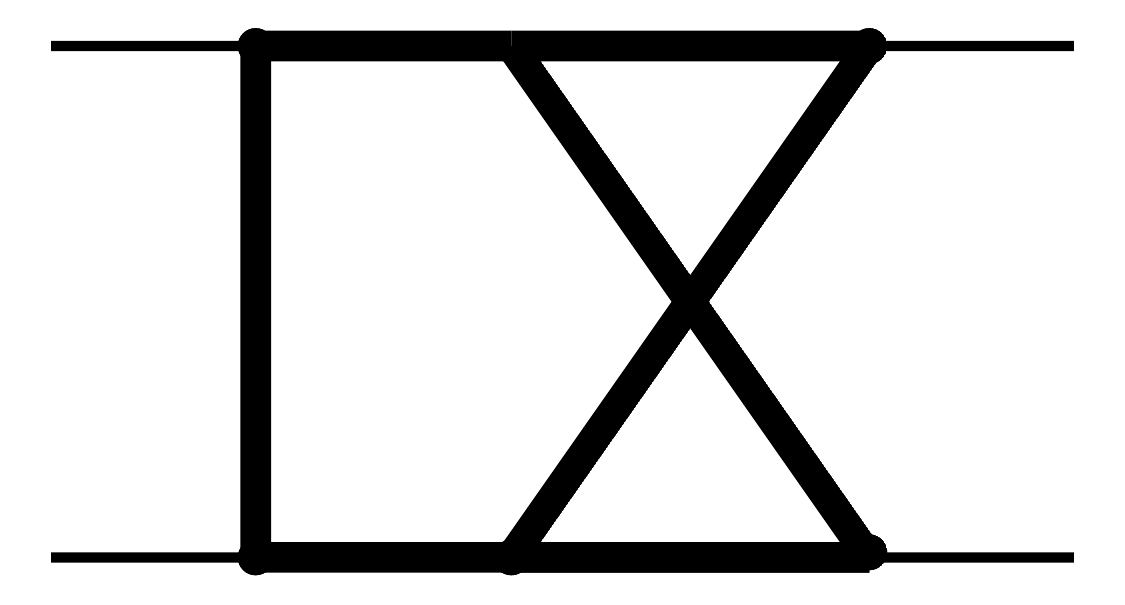}  \\
    (a) planar integral $\mathrm{PL}$ \qquad & \qquad (b) non-planar integral $\mathrm{NPL}$
    \end{tabular}
  \caption{\label{fig:FeynDiag}
   Diagrams for massive two-loop four-point Feynman integrals. Massive lines are thick ones.
    }
\end{figure}

There has been a long history of studying the high-energy behaviour in quantum field theory since V.V.~Sudakov in 1954~\cite{Sudakov:1954sw} 
and S.~Weinberg in 1960~\cite{Weinberg:1959nj}.  
In recent years, the study of massive scattering amplitudes in this limit at higher perturbative orders has drawn attention from the theoretical particle physics community, motivated by the need for high-precision predictions for experiments at the Large Hadron Collider~(LHC) at CERN.
The state of the art has advanced in several directions, including multi-loop and multi-leg massive QCD and QED amplitudes~\cite{Liu:2017vkm,Davies:2018ood,Davies:2018qvx,Mishima:2018olh,Catani:2022mfv,Buonocore:2023ljm,Wang:2023qbf,Wang:2024pmv,Lee:2024dbp,Lee:2024jvd,Jaskiewicz:2024xkd,Hu:2025aeo,Hu:2025hfc,Wang:2025kpk,Delto:2025epy} as well as electroweak amplitudes in the Standard Model~\cite{Davies:2022ram,Davies:2025wke,Zhang:2024frb}.
Notably, the calculation in the opposite (low-energy) limit has revealed interesting electroweak phenomenology for Higgs production processes~\cite{Davies:2023npk}.
Among the electroweak calculations, \texttt{AsyInt} has played an important role in solving two-loop master integrals for double Higgs production at high energies~\cite{Davies:2022ram,Davies:2025wke}.
Its mathematical aspects and technical details have been discussed in Ref.~\cite{Zhang:2024fcu}, and its repository is publicly available at
\[
\texttt{\href{https://gitlab.com/asyint/asyint-public}{https://gitlab.com/asyint/asyint-public}}
\]
with working examples.
In this proceedings, we present recent analytic results in the high-energy limit for massive two-loop four-point integrals shown in Fig.~\ref{fig:FeynDiag} as two examples.

The first simpler example is the planar double-box integral ($\mathrm{PL}$) with equal-mass internal and external lines shown in Fig.~\ref{fig:FeynDiag} (a).
The propagators of this integral are
\bea
&\Big\{-\left(l_1+q_1\right){}^2,m^2-\left(l_1+q_1+q_2\right){}^2,m^2-\left(l_2+q_1+q_2\right){}^2,-\left(l_2-q_3\right){}^2, \\ \nonumber
& m^2-l_2^2,m^2-l_1^2,-\left(l_2-l_1\right){}^2\Big\} \,,
\eea
where $l_1$ and $l_2$ are loop momenta and the external kinematics is 
\bea
q_1^2 = q_2^2 = q_3^2 = q_4^2 = m^2 \,, \quad  (q_1+q_2)^2 = s \,, \quad (q_1+q_3)^2 = t \,.
\eea
We apply the method of regions~\cite{Beneke:1997zp} 
by treating the mass as a small expansion parameter and find 17 relevant regions for this integral. We note that there is no so-called hidden region for four-point integrals at two loops~\cite{Gardi:2024axt}.
Using \texttt{AsyInt}, we obtain the Mellin-Barnes representations and solve them in terms of harmonic polylogarithms~\cite{Remiddi:1999ew} in the high-energy limit: $s, |t| \gg m^2$.
We express the final analytic result in the physical region ($s > 0$ and $T=-t>0$)
\bea
\calI_{\rm PL} & = &
\lb \frac{\mu^2}{s} \rb^{2 \epsilon}  \sum_{i=-2}^{0}  \, \eps^{i}  \, f^{(i)}\big(s,T,m^2 \big)\,,
\eea
with coefficient functions
\bea
f^{(-2)} & =&
\frac{1}{s^2 T}
 \Big[-2 \pi ^2+4 i \pi  H_0\big(
        \hat{m}^2
\big)
+2 H_0\big(
        \hat{m}^2\big)^2\Big]  \,, \\
f^{(-1)} & =&
\frac{1}{s^2 T} \bigg[
        \frac{i \pi ^3}{3}
        +4 \pi ^2 H_0\big(
                \hat{T}\big)
        -\frac{2}{3} H_0\big(
                \hat{m}^2\big)^3
        -2 i H_0\big(
                \hat{m}^2\big)^2 \big(
                \pi -2 i H_0\big(
                        \hat{T}\big)\big)
        \nonumber \\
        &&+\frac{1}{3} \pi  H_0\big(
                \hat{m}^2
        \big)
\big(5 \pi -24 i H_0\big(
                        \hat{T}\big)\big)
        -2 \zeta (3)
\bigg] \,, \\
f^{(0)} &=&
\frac{1}{s^2 T} \bigg[
        \frac{\pi ^4}{36}
        +\big(
                2 i \pi ^3
                -4 \pi ^2 H_1\big(
                        \hat{T}\big)
                -4 i \pi  H_{0,1}\big(
                        \hat{T}\big)
                +4 \zeta (3)
        \big) H_0\big(
                \hat{T}\big)
        \nonumber\\
        &&+H_0\big(
                \hat{m}^2\big)^2 \big(
                -\frac{8 \pi ^2}{3}+6 i \pi  H_0\big(
                        \hat{T}
                \big)
+2 H_0\big(
                        \hat{T}\big)^2\big)
        +\frac{2}{3} i \pi  H_0\big(
                \hat{T}\big)^3
                \nonumber\\
                && +\big(-\frac{8 i \pi ^3}{3}
                +\big(
                        -\frac{2 \pi ^2}{3}
                        +4 i \pi  H_1\big(
                                \hat{T}\big)
                        -4 H_{0,1}\big(
                                \hat{T}\big)
                \big) H_0\big(
                        \hat{T}\big)
                +\frac{2}{3} H_0\big(
                        \hat{T}\big)^3
                \nonumber\\
                &&-4 i \pi  H_{0,1}\big(
                        \hat{T}\big)
                +4 H_{0,0,1}\big(
                        \hat{T}\big)
                +2 H_0\big(
                        \hat{T}\big)^2 \big(
                        3 i \pi +H_1\big(
                                \hat{T}\big)\big)
                \nonumber\\
                &&+10 \zeta (3)
        \big) H_0\big(
                \hat{m}^2\big)
        -\frac{2}{3} H_0\big(
                \hat{m}^2\big)^4
        +4 \pi ^2 H_{0,1}\big(
                \hat{T}\big)
        +4 i \pi  H_{0,0,1}\big(
                \hat{T}\big)
        \nonumber\\
        &&+\frac{2}{3} H_0\big(
                \hat{m}^2\big)^3 \big(
                -i \pi +5 H_0\big(
                        \hat{T}\big)\big)
        -2 \pi  H_0\big(
                \hat{T}\big)^2 \big(
                2 \pi -i H_1\big(
                        \hat{T}\big)\big)
        +10 i \pi  \zeta (3)
\bigg] \,,
\eea
where $\eps$ is the dimensional regulator, $H_{\vec{w}} (z)$ are harmonic polylogarithms with $H_0(z) = \log(z)$, $\zeta(z)$ is the Riemann zeta function, and $\hat{T} = T/s$, $\hat{m}^2 = m^2/s$.
%
We anticipate this result will also be useful for studying the local two-loop IR subtractions~\cite{Anastasiou:2018rib,Ma:2019hjq} and rational terms~\cite{Pozzorini:2020hkx,Lang:2020nnl,Lang:2021hnw,Zhang:2022rft}.

%
The second more complicated example is the non-planar integral ($\mathrm{NPL}$) with fully massive internal lines shown in Fig.~\ref{fig:FeynDiag} (b). The propagators for this integral are
\bea
&\Big\{
m^2-l_1^2,m^2-\left(l_1+q_3\right){}^2,m^2-\left(l_1+l_2+q_2+q_3\right){}^2,m^2-\left(l_1+l_2-q_1\right){}^2, \nonumber \\
& m^2-\left(l_2-q_1\right){}^2,  m^2-l_2^2,m^2-\left(l_2+q_2\right){}^2
\Big\} \,,
\eea
where the external kinematics is
\bea
q_1^2 = q_2^2 = q_3^2 = q_4^2 = 0 \,, \quad  (q_1+q_2)^2 = s \,, \quad (q_1+q_3)^2 = t \,, \quad (q_2+q_3)^2 = u \,.
\eea
By assuming the internal mass to be small, we find 18 relevant regions whose calculations are highly non-trivial.
In particular, mathematical and technical complexities for this integral grow drastically with the increasing order of $\eps$ and $m$ in the dimensional regularisation and the high-energy expansion.
The first two leading terms in the high-energy limit up to $\ord(\eps^0)$ have been calculated in Ref.~\cite{Zhang:2024fcu} with \texttt{AsyInt},
and the more complete results to higher orders in $\eps$ and $m$ are calculated in Ref.~\cite{Davies:2025wke} with helps of differential equations and the analytic \texttt{Expand\&Fit} module of \texttt{AsyInt}.
Here we present the first two leading high-energy terms up to $\ord(\eps^0)$ in the physical region 
\bea
\calI_{\rm NPL}  & = &
\lb \frac{\mu^2}{s} \rb^{2 \epsilon}  \sum_{j=-1}^{0}  \, m^{j}  \, f_{(j)}\big(s,T,U,m^2 \big)\,,
\eea
with coefficient functions
\bea
f_{(-1)} 
&=&-\frac{i \, c_Z \, \pi^2 }{s^2 \sqrt{T U}}
 \,, \\[4pt]
 f_{(0)} 
 &=&
\frac{1}{s^2 T U} \bigg[
        \frac{ \pi ^4}{180} (15 s
        +202 T
        )
        -2 i \pi  \big(
                T \big(
                        \pi ^2-2 \zeta (3)\big)
                +s (-24+23 \zeta (3))
        \big)
        +\big(
                48 T
                -6 \pi ^2 T
                \nonumber\\
                &&+i \big(
                        24 \pi  U
                        +\frac{2}{3} \pi ^3 (7 s
                        -5 T
                        )
                \big)
                +\big(
                        12 i \pi  U
                        +\frac{1}{3} \big(
                                -\big(
                                        \big(
                                                72+13 \pi ^2\big) s\big)
                                +14 \pi ^2 T
                        \big)
                        \nonumber\\
                        &&+6 (T
                        -U
                        ) H_{0,1}\big(
                                \hat{T}\big)
                \big) H_1\big(
                        \hat{T}\big)
                +(6+i \pi ) s H_1\big(
                        \hat{T}\big)^2
                +\frac{1}{3} (s
                -4 T
                ) H_1\big(
                        \hat{T}\big)^3
                \nonumber\\
                &&+(12 (T-2 s
                )
                +4 i \pi  (2 s
                +T
                )
                ) H_{0,1}\big(
                        \hat{T}\big)
                +(24 T -32 s
                ) H_{0,0,1}\big(
                        \hat{T}\big)
                +14 (U-T
                ) H_{0,1,1}\big(
                        \hat{T}\big)
                \nonumber\\
                &&+\frac{2}{3} (s
                -34 T
                ) \zeta (3)
        \big) H_0\big(
                \hat{T}\big)
        +\big(
                6 i \pi  U
                +\frac{1}{3} \pi ^2 (-3 s
                +7 T
                )
                +(-2 i \pi  T
                +6 U
                ) H_1\big(
                        \hat{T}\big)
                \nonumber\\
                &&-T H_1\big(
                        \hat{T}\big)^2
                +(8 s
                -6 T
                ) H_{0,1}\big(
                        \hat{T}\big)
        \big) H_0\big(
                \hat{T}\big)^2
        +\big(
                -2 T
                +\frac{1}{3} i \pi  (5 s
                -2 T
                )
                \nonumber\\
                &&+\big(
                        -s
                        +\frac{4 T}{3}
                \big) H_1\big(
                        \hat{T}
                \big)
        \big) H_0\big(
                \hat{T}\big)^3
        +
        \frac{1}{6} (s
        +2 T
        ) H_0\big(
                \hat{T}\big)^4
        +\big(
                -\frac{2}{3} i \pi  \big(
                        36+7 \pi ^2\big) s
                        \nonumber\\
                        &&+\big( -12 i \pi  U
                        -\frac{8}{3} \big(
                                \big(
                                        9+\pi ^2\big) T
                                -\pi ^2 U
                        \big)
                        +2 (6+i \pi ) s H_1\big(
                                \hat{T}\big)
                        +2 U H_1\big(
                                \hat{T}\big)^2
                \big) H_0\big(
                        \hat{T}\big)
                \nonumber\\
                &&+\big(
                        -2 i \pi  s
                        -2 T H_1\big(
                                \hat{T}\big)
                \big) H_0\big(
                        \hat{T}\big)^2
                -\frac{2}{3} T H_0\big(
                        \hat{T}\big)^3
                \nonumber\\
                &&+\big(
                        12 i \pi  T
                        +24 U
                        +\frac{8}{3} \pi ^2 (-T
                        +U
                        )
                \big) H_1\big(
                        \hat{T}\big)
                -2 i \pi  s H_1\big(
                        \hat{T}\big)^2
                +\frac{2}{3} U H_1\big(
                        \hat{T}\big)^3
                \nonumber\\
                &&-\frac{8 s \zeta (3)}{3}
        \big) H_0\big(
                \hat{m}^2\big)
        +\big(
                6 i \pi  s
                -\frac{5 \pi ^2 s}{3}
                +\big(
                        6 T
                        +i \pi  (T
                        -U
                        )
                        +3 s H_1\big(
                                \hat{T}\big)
                \big) H_0\big(
                        \hat{T}\big)
                \nonumber\\
                &&-s H_0\big(
                        \hat{T}\big)^2
                +(-6 U
                +i \pi  (T
                -U
                )
                ) H_1\big(
                        \hat{T}\big)
                -s H_1\big(
                        \hat{T}\big)^2
        \big) H_0\big(
                \hat{m}^2\big)^2
                \nonumber\\
                && +\big( \frac{4 i \pi  s}{3}
                +\frac{2}{3} (2 s
                +T
                ) H_0\big(
                        \hat{T}\big)
                +\big(
                        -2 s
                        +\frac{2 T}{3}
                \big) H_1\big(
                        \hat{T}
                \big)
        \big) H_0\big(
                \hat{m}^2\big)^3
        -
        \frac{1}{2} s H_0\big(
                \hat{m}^2\big)^4
        \nonumber\\
        &&+\big(
                6 \big(
                        \pi ^2 -8 \big) U
                +i \big(
                        -24 \pi  T
                        +\frac{\pi ^3}{3}  (s
                        -10 T
                        )
                \big)
                -10 i \pi  s H_{0,1}\big(
                        \hat{T}\big)
                +6 (U-T
                ) H_{0,0,1}\big(
                        \hat{T}\big)
                \nonumber\\
                &&+4 (s
                +3 T
                ) H_{0,1,1}\big(
                        \hat{T}\big)
                +\frac{4}{3} (9 s
                -17 T
                ) \zeta (3)
        \big) H_1\big(
                \hat{T}\big)
        +\big(
                6 i \pi  T
                \nonumber\\
                &&+\frac{1}{3} \pi ^2 (4 s
                -7 T
                )
        \big) H_1\big(
                \hat{T}\big)^2
        +\big(
                2 U
                -\frac{1}{3} i \pi  (3 s
                +2 T
                )
        \big) H_1\big(
                \hat{T}\big)^3
        +\frac{1}{6} (3 s
        -2 T
        ) H_1\big(
                \hat{T}\big)^4
        \nonumber\\
        &&+\big(
                12 i \pi  (T
                -U
                )
                +\frac{14}{3} \pi ^2 (
                U-T
                )
        \big) H_{0,1}\big(
                \hat{T}\big)
        +(24 s
        -12 T
        -2 i \pi  (9 s
        +2 T
        )
        ) H_{0,0,1}\big(
                \hat{T}\big)
        \nonumber\\
        &&+(i \pi  (22 s
        -4 T
        )
        -12 (s
        +T
        )
        ) H_{0,1,1}\big(
                \hat{T}\big)
        +(48 s
        -36 T
        ) H_{0,0,0,1}\big(
                \hat{T}\big)
        \nonumber\\
        &&+20 (T
        -U
        ) H_{0,0,1,1}\big(
                \hat{T}\big)
        -12 (s
        +3 T
        ) H_{0,1,1,1}\big(
                \hat{T}\big)
        +\frac{4}{9} \pi ^2 s \psi ^{(1)}\big(
                \frac{1}{3}\big)
        -\frac{1}{3} s \psi ^{(1)}\big(
                \frac{1}{3}\big)^2
        \nonumber\\
        &&+12 T \zeta (3)
\bigg] \,,
\eea
where $\psi^{(i)}(z)$ is the polygamma function, $\hat{T} = T/s$, $\hat{m}^2 = m^2/s$ and  $U = s - T$.
Note that a series definition of the constant $c_Z$ is found in Ref.~\cite{Zhang:2024fcu}, 
and its closed form is found in Ref.~\cite{Davies:2025wke}
\bea
c_Z
&  = & \int_{0}^{\infty} \int_{0}^{\infty} \, \frac{\mathrm{d} \alpha_1 \, \mathrm{d}  \alpha_2}{\sqrt{\alpha _1 \, \alpha _2 \, \big( \alpha _1+\alpha _2+1\big) \, \big( \alpha _2 \alpha _1+\alpha _1+\alpha _2\big) } }
 \; = \; 4 \sqrt{3} \, K^2\Big(\frac{1}{2}-\frac{\sqrt{3}}{4}\Big) \,,
\eea 
where  $K(z)$ is the complete elliptic function of the first kind. 
Note that at higher expansion orders another constant is found in Ref.~\cite{Davies:2025wke}
\bea
        c_{Z_2} 
                & = & \Big(1+\frac{1}{\sqrt{3}}\Big) \pi + \Big(2+\frac{17}{4 \sqrt{3}}\Big) K^2\Big(\frac{1}{2}-\frac{\sqrt{3}}{4}\Big) -4 \sqrt{3} E^2\Big(\frac{1}{2}-\frac{\sqrt{3}}{4}\Big)  \, ,
\eea
with $E(z)$ being the complete elliptic function of the second kind.
We further find a relation
\bea
\pi &=& -2 \, K\Big(\frac{1}{2}-\frac{\sqrt{3}}{4}\Big) \left[\left(\sqrt{3}+1\right) K\Big(\frac{1}{2}-\frac{\sqrt{3}}{4}\Big)-2 \sqrt{3} E\Big(\frac{1}{2}-\frac{\sqrt{3}}{4}\Big)\right] \,,
\eea
which indicates a half-integer transcendental weight for $K\Big(\frac{1}{2}-\frac{\sqrt{3}}{4}\Big)$ and $E\Big(\frac{1}{2}-\frac{\sqrt{3}}{4}\Big)$.
The deep high-energy expansion of this integral has been successfully applied to next-to-leading-order two-loop Yukawa and Higgs self-coupling corrections to double Higgs production~\cite{Davies:2025wke}, and the analytic results are in good agreement with numerical results in Ref.~\cite{Heinrich:2024dnz}.

\section*{Acknowledgment}
The author thanks Joshua Davies, Yao Ma, Kay Sch\"onwald, Matthias Steinhauser for useful discussions. This research is funded by the European Union under the Marie Sk{\l}odowska-Curie Actions (MSCA) grant 101202083 -- ``HINOVA".

\small
\setlength{\bibsep}{3pt}
\bibliographystyle{JHEP}
\bibliography{reference}

\providecommand{\href}[2]{#2}\begingroup\raggedright\begin{thebibliography}{10}

\bibitem{Zhang:2024fcu}
H.~Zhang, \emph{{Massive two-loop four-point Feynman integrals at high energies
  with AsyInt}}, \href{https://doi.org/10.1007/JHEP09(2024)069}{\emph{JHEP}
  {\bfseries 09} (2024) 069}
  [\href{https://arxiv.org/abs/2407.12107}{{\ttfamily 2407.12107}}].

\bibitem{Sudakov:1954sw}
V.~V. Sudakov, \emph{{Vertex Parts at Very High Energies in Quantum
  Electrodynamics}}, {\emph{Sov. Phys. JETP} {\bfseries 3,4,5,6} (1956) 65}.

\bibitem{Weinberg:1959nj}
S.~Weinberg, \emph{{High-energy behavior in quantum field theory}},
  \href{https://doi.org/10.1103/PhysRev.118.838}{\emph{Phys. Rev.} {\bfseries
  118} (1960) 838}.

\bibitem{Liu:2017vkm}
T.~Liu and A.~A. Penin, \emph{{High-Energy Limit of QCD beyond the Sudakov
  Approximation}},
  \href{https://doi.org/10.1103/PhysRevLett.119.262001}{\emph{Phys. Rev. Lett.}
  {\bfseries 119} (2017) 262001}
  [\href{https://arxiv.org/abs/1709.01092}{{\ttfamily 1709.01092}}].

\bibitem{Davies:2018ood}
J.~Davies, G.~Mishima, M.~Steinhauser and D.~Wellmann, \emph{{Double-Higgs
  boson production in the high-energy limit: planar master integrals}},
  \href{https://doi.org/10.1007/JHEP03(2018)048}{\emph{JHEP} {\bfseries 03}
  (2018) 048} [\href{https://arxiv.org/abs/1801.09696}{{\ttfamily
  1801.09696}}].

\bibitem{Davies:2018qvx}
J.~Davies, G.~Mishima, M.~Steinhauser and D.~Wellmann, \emph{{Double Higgs
  boson production at NLO in the high-energy limit: complete analytic
  results}}, \href{https://doi.org/10.1007/JHEP01(2019)176}{\emph{JHEP}
  {\bfseries 01} (2019) 176}
  [\href{https://arxiv.org/abs/1811.05489}{{\ttfamily 1811.05489}}].

\bibitem{Mishima:2018olh}
G.~Mishima, \emph{{High-Energy Expansion of Two-Loop Massive Four-Point
  Diagrams}}, \href{https://doi.org/10.1007/JHEP02(2019)080}{\emph{JHEP}
  {\bfseries 02} (2019) 080}
  [\href{https://arxiv.org/abs/1812.04373}{{\ttfamily 1812.04373}}].

\bibitem{Catani:2022mfv}
S.~Catani, S.~Devoto, M.~Grazzini, S.~Kallweit, J.~Mazzitelli and C.~Savoini,
  \emph{{Higgs Boson Production in Association with a Top-Antitop Quark Pair in
  Next-to-Next-to-Leading Order QCD}},
  \href{https://doi.org/10.1103/PhysRevLett.130.111902}{\emph{Phys. Rev. Lett.}
  {\bfseries 130} (2023) 111902}
  [\href{https://arxiv.org/abs/2210.07846}{{\ttfamily 2210.07846}}].

\bibitem{Buonocore:2023ljm}
L.~Buonocore, S.~Devoto, M.~Grazzini, S.~Kallweit, J.~Mazzitelli, L.~Rottoli
  et~al., \emph{{Precise Predictions for the Associated Production of a W Boson
  with a Top-Antitop Quark Pair at the LHC}},
  \href{https://doi.org/10.1103/PhysRevLett.131.231901}{\emph{Phys. Rev. Lett.}
  {\bfseries 131} (2023) 231901}
  [\href{https://arxiv.org/abs/2306.16311}{{\ttfamily 2306.16311}}].

\bibitem{Wang:2023qbf}
G.~Wang, T.~Xia, L.~L. Yang and X.~Ye, \emph{{On the high-energy behavior of
  massive QCD amplitudes}},
  \href{https://doi.org/10.1007/JHEP05(2024)082}{\emph{JHEP} {\bfseries 05}
  (2024) 082} [\href{https://arxiv.org/abs/2312.12242}{{\ttfamily
  2312.12242}}].

\bibitem{Wang:2024pmv}
G.~Wang, T.~Xia, L.~L. Yang and X.~Ye, \emph{{Two-loop QCD amplitudes for $
  t\overline{t}H $ production from boosted limit}},
  \href{https://doi.org/10.1007/JHEP07(2024)121}{\emph{JHEP} {\bfseries 07}
  (2024) 121} [\href{https://arxiv.org/abs/2402.00431}{{\ttfamily
  2402.00431}}].

\bibitem{Lee:2024dbp}
R.~N. Lee and V.~A. Stotsky, \emph{{Master integrals for e$^{+}$e$^{-}$
  {\textrightarrow} 2{\ensuremath{\gamma}} process at large energies and
  angles}}, \href{https://doi.org/10.1007/JHEP12(2024)106}{\emph{JHEP}
  {\bfseries 12} (2024) 106}
  [\href{https://arxiv.org/abs/2410.03336}{{\ttfamily 2410.03336}}].

\bibitem{Lee:2024jvd}
R.~N. Lee, \emph{{Two-loop master integrals for e$^{+}$e$^{-}${\textrightarrow}
  {\ensuremath{\mu}}$^{+}${\ensuremath{\mu}}$^{-}$ process with account of
  electron mass}}, \href{https://doi.org/10.1007/JHEP02(2025)006}{\emph{JHEP}
  {\bfseries 02} (2025) 006}
  [\href{https://arxiv.org/abs/2412.00793}{{\ttfamily 2412.00793}}].

\bibitem{Jaskiewicz:2024xkd}
S.~Jaskiewicz, S.~Jones, R.~Szafron and Y.~Ulrich, \emph{{The structure of
  quark mass corrections in the gg {\textrightarrow} HH amplitude at
  high-energy}}, \href{https://doi.org/10.1007/JHEP09(2025)015}{\emph{JHEP}
  {\bfseries 09} (2025) 015}
  [\href{https://arxiv.org/abs/2501.00587}{{\ttfamily 2501.00587}}].

\bibitem{Hu:2025aeo}
Z.~Hu, T.~Liu and J.~M. Yang, \emph{{The gg {\textrightarrow} HH amplitude
  induced by bottom quarks at two-loop level: planar master integrals}},
  \href{https://doi.org/10.1007/JHEP09(2025)132}{\emph{JHEP} {\bfseries 09}
  (2025) 132} [\href{https://arxiv.org/abs/2503.10051}{{\ttfamily
  2503.10051}}].

\bibitem{Hu:2025hfc}
Z.~Hu and T.~Liu, \emph{{Double logarithmic contribution to Higgs pair
  production in the high-energy limit}},
  \href{https://arxiv.org/abs/2509.06381}{{\ttfamily 2509.06381}}.

\bibitem{Wang:2025kpk}
G.~Wang and L.~L. Yang, \emph{{Two-loop QCD amplitudes for $t\bar{t}\gamma$
  production at hadron colliders}},
  \href{https://arxiv.org/abs/2510.01774}{{\ttfamily 2510.01774}}.

\bibitem{Delto:2025epy}
M.~Delto, A.~Penin and L.~Tancredi, \emph{{High-Energy Evolution of
  Power-Suppressed Amplitudes}},
  \href{https://arxiv.org/abs/2510.04914}{{\ttfamily 2510.04914}}.

\bibitem{Davies:2022ram}
J.~Davies, G.~Mishima, K.~Sch\"onwald, M.~Steinhauser and H.~Zhang,
  \emph{{Higgs boson contribution to the leading two-loop Yukawa corrections to
  gg \textrightarrow{} HH}},
  \href{https://doi.org/10.1007/JHEP08(2022)259}{\emph{JHEP} {\bfseries 08}
  (2022) 259} [\href{https://arxiv.org/abs/2207.02587}{{\ttfamily
  2207.02587}}].

\bibitem{Davies:2025wke}
J.~Davies, K.~Sch{\"o}nwald, M.~Steinhauser and H.~Zhang, \emph{{Analytic
  next-to-leading order Yukawa and Higgs boson self-coupling corrections to gg
  {\textrightarrow} HH at high energies}},
  \href{https://doi.org/10.1007/JHEP04(2025)193}{\emph{JHEP} {\bfseries 04}
  (2025) 193} [\href{https://arxiv.org/abs/2501.17920}{{\ttfamily
  2501.17920}}].

\bibitem{Zhang:2024frb}
H.~Zhang, \emph{{QCD and electroweak corrections for single and double Higgs
  boson production at the LHC}},
  \href{https://doi.org/10.22323/1.476.0074}{\emph{PoS} {\bfseries ICHEP2024}
  (2025) 074} [\href{https://arxiv.org/abs/2409.15438}{{\ttfamily
  2409.15438}}].

\bibitem{Davies:2023npk}
J.~Davies, K.~Sch\"onwald, M.~Steinhauser and H.~Zhang, \emph{{Next-to-leading
  order electroweak corrections to $gg \to HH$ and $gg \to gH$ in the
  large-$m_t$ limit}},
  \href{https://doi.org/10.1007/JHEP10(2023)033}{\emph{JHEP} {\bfseries 10}
  (2023) 033} [\href{https://arxiv.org/abs/2308.01355}{{\ttfamily
  2308.01355}}].

\bibitem{Beneke:1997zp}
M.~Beneke and V.~A. Smirnov, \emph{{Asymptotic expansion of Feynman integrals
  near threshold}},
  \href{https://doi.org/10.1016/S0550-3213(98)00138-2}{\emph{Nucl. Phys. B}
  {\bfseries 522} (1998) 321}
  [\href{https://arxiv.org/abs/hep-ph/9711391}{{\ttfamily hep-ph/9711391}}].

\bibitem{Gardi:2024axt}
E.~Gardi, F.~Herzog, S.~Jones and Y.~Ma, \emph{{Dissecting polytopes: Landau
  singularities and asymptotic expansions in 2 {\textrightarrow} 2
  scattering}}, \href{https://doi.org/10.1007/JHEP08(2024)127}{\emph{JHEP}
  {\bfseries 08} (2024) 127}
  [\href{https://arxiv.org/abs/2407.13738}{{\ttfamily 2407.13738}}].

\bibitem{Remiddi:1999ew}
E.~Remiddi and J.~A.~M. Vermaseren, \emph{{Harmonic polylogarithms}},
  \href{https://doi.org/10.1142/S0217751X00000367}{\emph{Int. J. Mod. Phys. A}
  {\bfseries 15} (2000) 725}
  [\href{https://arxiv.org/abs/hep-ph/9905237}{{\ttfamily hep-ph/9905237}}].

\bibitem{Anastasiou:2018rib}
C.~Anastasiou and G.~Sterman, \emph{{Removing infrared divergences from
  two-loop integrals}},
  \href{https://doi.org/10.1007/JHEP07(2019)056}{\emph{JHEP} {\bfseries 07}
  (2019) 056} [\href{https://arxiv.org/abs/1812.03753}{{\ttfamily
  1812.03753}}].

\bibitem{Ma:2019hjq}
Y.~Ma, \emph{{A Forest Formula to Subtract Infrared Singularities in Amplitudes
  for Wide-angle Scattering}},
  \href{https://doi.org/10.1007/JHEP05(2020)012}{\emph{JHEP} {\bfseries 05}
  (2020) 012} [\href{https://arxiv.org/abs/1910.11304}{{\ttfamily
  1910.11304}}].

\bibitem{Pozzorini:2020hkx}
S.~Pozzorini, H.~Zhang and M.~F. Zoller, \emph{{Rational Terms of UV Origin at
  Two Loops}}, \href{https://doi.org/10.1007/JHEP05(2020)077}{\emph{JHEP}
  {\bfseries 05} (2020) 077}
  [\href{https://arxiv.org/abs/2001.11388}{{\ttfamily 2001.11388}}].

\bibitem{Lang:2020nnl}
J.-N. Lang, S.~Pozzorini, H.~Zhang and M.~F. Zoller, \emph{{Two-Loop Rational
  Terms in Yang-Mills Theories}},
  \href{https://doi.org/10.1007/JHEP10(2020)016}{\emph{JHEP} {\bfseries 10}
  (2020) 016} [\href{https://arxiv.org/abs/2007.03713}{{\ttfamily
  2007.03713}}].

\bibitem{Lang:2021hnw}
J.-N. Lang, S.~Pozzorini, H.~Zhang and M.~F. Zoller, \emph{{Two-loop rational
  terms for spontaneously broken theories}},
  \href{https://doi.org/10.1007/JHEP01(2022)105}{\emph{JHEP} {\bfseries 01}
  (2022) 105} [\href{https://arxiv.org/abs/2107.10288}{{\ttfamily
  2107.10288}}].

\bibitem{Zhang:2022rft}
H.~Zhang, \emph{{UV and IR rational terms in two-loop amplitudes: first
  insights}}, \href{https://doi.org/10.22323/1.416.0072}{\emph{PoS} {\bfseries
  LL2022} (2022) 072}.

\bibitem{Heinrich:2024dnz}
G.~Heinrich, S.~Jones, M.~Kerner, T.~Stone and A.~Vestner, \emph{{Electroweak
  corrections to Higgs boson pair production: the top-Yukawa and self-coupling
  contributions}}, \href{https://doi.org/10.1007/JHEP11(2024)040}{\emph{JHEP}
  {\bfseries 11} (2024) 040}
  [\href{https://arxiv.org/abs/2407.04653}{{\ttfamily 2407.04653}}].

\end{thebibliography}\endgroup

%

\end{document}